\documentclass[preprint2]{aa}

\usepackage{graphicx}
\begin{document}
\def\P{\bar{\Phi}}

\def\st{\sigma_{\rm T}}

\def\vk{v_{\rm K}}

\def\sles{\lower2pt\hbox{$\buildrel {\scriptstyle <}
   \over {\scriptstyle\sim}$}}

\def\sgreat{\lower2pt\hbox{$\buildrel {\scriptstyle >}
   \over {\scriptstyle\sim}$}}

\title{Three-dimensional numerical simulations of the pulsar magnetosphere:
Preliminary results}

\author{Constantinos Kalapotharakos and Ioannis Contopoulos}
\institute{Research Center for Astronomy, Academy of Athens,
GR-11527 Athens, Greece\\
\email{ckalapot@phys.uoa.gr,\ icontop@academyofathens.gr}}

\titlerunning{3D Pulsar}

\date{Received 29 May 2008/ Accepted 14 November 2008}

\abstract {We investigate the three-dimensional structure of the
pulsar magnetosphere through time-dependent numerical simulations
of a magnetic dipole that is set in rotation. We developed our own
Eulerian finite difference time domain numerical solver of
force-free electrodynamics and implemented the technique of
non-reflecting and absorbing outer boundaries. This allows us to
run our simulations for many stellar rotations, and thus claim
with confidence that we have reached a steady state. A
quasi-stationary corotating pattern is established, in agreement
with previous numerical solutions. We discuss the prospects of our
code for future high-resolution investigations of dissipation,
particle acceleration, and temporal variability.

}

\maketitle
\section{Introduction}

In the 40 years following the discovery of pulsars, significant
progress has been made towards understanding the pulsar phenomenon
(e.g. Michel~1991; Bass {\em et al.}~2008). We know that pulsars
are magnetized neutron stars with non-aligned rotation and
magnetic axes (oblique rotators). We also know that pulsars lose
rotational energy and spin down through electromagnetic torques
due to large-scale electric currents in their magnetospheres.

Unfortunately, several pieces of the puzzle are still missing. In
particular, we have only a vague idea about the structure of the
pulsar magnetosphere. An analytic expression only exists for the
structure of the magnetosphere of the vacuum rotator (the retarded
dipole solution, Deutsch~1955), but real pulsars are certainly not
in vacuum since electrons and positrons are copiously produced due
to the high surface electric fields induced by rotation. Numerical
solutions are hard to obtain because the magnetosphere develops
singular current sheets, and the problem is fundamentally
three-dimensional with an extended spatial dynamic range. Our
current understanding is based on the axisymmetric solution
obtained for the first time in Contopoulos, Kazanas \&
Fendt~(1999) (hereafter CKF). This solution has since been
confirmed, improved, and generalized by several other authors
(e.g. Gruzinov~2005, Contopoulos~2005, Timokhin~2006,
Komissarov~2006, McKinney~2006). The first time-dependent
numerical simulations of the pulsar magnetosphere in 3D were
performed by Spitkovsky~(2006).

Our current understanding suggests that the general 3D
magnetosphere consists of regions of closed and open field lines
(those that are stretched out to infinity away from their point of
origin on the surface of the neutron star). It also suggests that
a large-scale electric current circuit is established along open
magnetic field lines. In that picture, electric current closure is
guaranteed through a current sheet that flows in the equatorial
region and along the boundary between open and closed field lines.
The equatorial current sheet is confined between latitudes
$\pm\theta$ above and below the rotation equator ($\theta$ being
the inclination angle between the rotation and magnetic axes) and
has an undulating shape of a spiral form with radial wavelength
equal to $2\pi$ times the light cylinder distance $r_{lc}\equiv
c/\Omega_*$, where $\Omega_*$ is the angular velocity of rotation
of the central neutron star (e.g. Bogovalov 1999). A large-scale
pattern of electromagnetic energy flow (Poynting flux) and charged
relativistic particle wind is established along open field lines.
The details of how the particles are supplied are not understood
well. Moreover, the equatorial current sheet is probably unstable,
and may not survive beyond the light cylinder (Romanova, Chulsky
\& Lovelace~2005). This is different from the stable heliospheric
equatorial current sheet that is simply the tangential
discontinuity convected with the velocity of the solar wind plasma
(Landau \& Lifshitz~1969).

The details of the 3D solution are of paramount importance for
answering several questions about the pulsar phenomenon such as
where in the magnetosphere the observed electromagnetic radiation
is coming from, what determines the radiation spectrum and the
pulse profile, what accelerates the pulsar wind, etc. Guided by
our experience from the solution of the steady-state axisymmetric
problem, we believe that a promising approach to obtaining the
structure of the 3D pulsar magnetosphere may be through time
dependent relativistic ideal MHD numerical simulations of a
rotating magnetized star which, when run long enough, will
hopefully relax to the steady-state solution.

The pulsar magnetosphere problem belongs to a subclass of
relativistic MHD, that of Force-Free Electrodynamics, hereafter
FFE (e.g. Gruzinov~1999). FFE assumes that the relativistic medium
is on the one hand dense enough to provide charge carriers that
will guarantee `infinite' plasma conductivity and therefore
\begin{equation}
{\bf E}\cdot {\bf B}=0\ \mbox{everywhere,}
\label{EperpB}
\end{equation}
and on the other hand tenuous enough for plasma inertia and pressure
terms to be neglected and therefore
\begin{equation}
\rho_e {\bf E} + {\bf J}\times {\bf B} = 0\
\mbox{everywhere.}
\label{forcefree}
\end{equation}
Here,
\begin{equation}
\rho_e\equiv \frac{1}{4\pi}\nabla\cdot {\bf E}
\end{equation}
is the charge density distributed in space, and ${\bf J}$ is the
electric current. The pulsar magnetosphere is a unique physical
system in which the above conditions are satisfied over its
largest part. Deviations from FFE do exist in singular regions of
the magnetosphere, and these introduce important physical
complications to the problem.

As we said above, the first FFE numerical simulations of the 3D
pulsar magnetosphere were performed by Spitkovsky~(2006).
Our main concern with those simulations has been that they run
only for a limited amount of time which may not have been enough
to reach steady state (see \S~2). Since no equivalent numerical
simulations existed up to now in the literature, we have tried to
reproduce Spitkovsky's conclusions based on our experience from
two idealized cases where we know the solution almost analytically
(Contopoulos~2007), but without success. In view of the above, and
because of the central role understanding the structure of the
pulsar magnetosphere plays for pulsar research, we decided to
independently develop our own Finite-Difference Time-Domain
(hereafter FDTD; Taflove \& Hagness 2005) FFE code. The new
element is that we have now implemented the technique of Perfectly
Matched Layer (hereafter PML; Berenger 1994, 1996) which is quite
efficient in minimizing reflection and maximizing absorbtion from
the outer boundaries of the simulation box. Such boundary
conditions imitate open space, allowing thus to run our
simulations for very long times up to satisfactory numerical
convergence to a steady state. As we will see below, we are
currently in a position to address the same problem, in comparable
detail, using just a standard off-the-shelf PC.

In \S~2 we justify our effort to implement PML outer boundary
conditions by laying out our criticism of existing numerical
results. In \S~3, we describe the computer code that we have
developed and the solutions to the numerical problems that we
encountered when running our simulations. Our first results and
discoveries with detail comparable to existing numerical
simulations are presented in \S~4. We conclude in \S~5 with a
discussion on numerical convergence and tests, as well as a short
presentation of future prospects for our code.

\section{Need for longer numerical simulations}

We will assume that relativistic ideal MHD, and in particular its
Force-Free idealization is a valid description of the pulsar
magnetosphere (we will get back to the issue of non-ideal
processes in \S~5). Therefore, the magnetospheric electric and
magnetic fields ${\bf E}$ and ${\bf B}$ satisfy not only Maxwell's
equations
\begin{equation}
\frac{\partial {\bf E}}{\partial t} = c\nabla\times {\bf B}
-4\pi{\bf J}\ ,
\label{Maxwell1}
\end{equation}
\begin{equation}
\frac{\partial {\bf B}}{\partial t} = -c\nabla\times {\bf E}\ ,\ \mbox{and}
\label{Maxwell2}
\end{equation}
\begin{equation}
\nabla\cdot {\bf B}=0\ ,
\label{Maxwell3}
\end{equation}
but also eqs.~(\ref{EperpB}) and (\ref{forcefree}). After some
algebra, we obtain
\begin{equation}
{\bf J}= \rho_e c\frac{{\bf E}\times {\bf B}}{B^2}+
\frac{c}{4\pi}\frac{({\bf B}\cdot \nabla\times {\bf B} -{\bf
E}\cdot \nabla\times {\bf E})}{B^2}{\bf B}
\label{J}
\end{equation}
(Gruzinov~1999, 2005). Here, $c{\bf E}\times {\bf B}/B^2$ is the
pulsar wind drift velocity. As we will see in the next section, a
stationary corotating pattern is established in the pulsar
magnetosphere, where the electric field is given by
\begin{equation}
{\bf E}=-\frac{1}{c}({\bf \Omega}_*\times {\bf r}) \times {\bf B}
\label{corotation}
\end{equation}
(e.g. Muslimov \& Harding~2005).

If one starts with a magnetic field configuration `anchored' onto
the rotating magnetized neutron star (e.g. a simple magnetostatic
dipole), and if one sets the star in rotation, electric currents
will develop which will populate the magnetosphere with electric
charge originating in the stellar surface. It is natural to expect
that if one integrates eqs.~(\ref{Maxwell1}) and (\ref{Maxwell2})
long enough, the magnetosphere will relax to the final steady
state, if such a steady state indeed exists. This will determine
the distribution of electric charge, electric current,
electromagnetic energy flow (Poynting flux), and pulsar wind drift
velocities in the pulsar magnetosphere.

As we said, there are some important complications in this
approach. The first one is that as the simulation evolves, current
sheets appear where ${\bf J}$ becomes formally infinite and FFE
breaks down. In the equatorial current sheet, the magnetic field
goes to zero. At those positions, the electric field risks
becoming greater than the magnetic field, and the drift velocity
risks becoming greater than the speed of light.
Physically, this corresponds to runaway particle acceleration (we
will come back to this point in \S~5), therefore, some way of
restricting drift velocities to subluminal values must be
incorporated in the code. The second complication is the
fundamental 3D nature of the problem. In order to treat both the
smallest (current sheets, neutron star polar cap) and largest
physical scales of the system (light cylinder, equatorial current
sheet undulations), one needs a numerical simulation with
sufficient spatial dynamic range. The simulation must run long
enough to reach convergence and come up with a believable
solution. Without the implementation of non-reflecting and
absorbing outer boundaries, simulation runs are limited by the
time it takes for the transient wave that results from the
initiation of the central neutron star rotation to reach the outer
boundary of the simulation and return to affect the region of
interest. It is precisely this complication that limits current
state of the art computer simulations to less than about 2 central
neutron star rotations, and the region of believable results not
much further than about 2 light cylinder radii. We are, therefore,
convinced that a promising approach to making significant progress
is to run the simulations for much longer timescales by
implementing non-reflecting and absorbing outer boundaries in the
code.

Another interesting issue has to do with the pulsar spindown rate
$L$ as a function of the inclination angle $\theta$ as obtained by
Spitkovsky~(2006), namely
\begin{equation}
L(\theta)= \frac{B_*^2 r_*^6 \Omega_*^4}{4c^3}(1+\sin^2\theta)\ ,
\label{Spitkovsky}
\end{equation}
where $B_*$, $r_*$ are the polar dipole magnetic field and radius
of the central neutron star respectively. Most people are content
with the approximation that pulsars spin down at the same rate as
$90^o$ vacuum dipole rotators, namely that $L(\theta)\approx B_*^2
r_*^6 \Omega_*^4/(6c^3)$; however, this is not accurate, since the
dependence of $L$ on the magnetic inclination angle $\theta$ has
important implications for the evolution and distribution of
pulsars in the $P-\dot{P}$ diagram (Contopoulos \&
Spitkovsky~2006). We would like to independently confirm
eq.~\ref{Spitkovsky} with our own numerical code.

\section{FFE with central symmetry and PML outer
boundaries}

In order to solve the system of eqs.~(\ref{Maxwell1}) and
(\ref{Maxwell2}) we implemented the FDTD Yee algorithm (Yee~1966).
According to this, electric field components are defined parallel
to the mesh cell sides, whereas magnetic field components are
defined perpendicular to mesh cell faces. Apparently, this
staggered mesh configuration guarantees low numerical diffusivity,
something that is required in electrodynamic problems with sharp
gradients such as the one we are presently addressing. The field
components required at other mesh positions are obtained by
first-order interpolation. We decided to implement a cartesian
numerical grid $(x,y,z) =(i\delta,j\delta,k\delta)$ (where $i,j,k$
are integers and $\delta$ is the grid spatial resolution) rather
than a cylindrical or spherical numerical one in order not to have
to deal with numerical problems around the rotation axis.
Moreover, instead of the staggered leapfrog time integration for
the electric and magnetic field components of the original Yee
algorithm, we chose to use third-order Runge-Kutta for the time
integration. This approach is more accurate and provides both the
electric and magnetic field at the same time moments. Finally,
because of the special nature of the pulsar magnetosphere problem,
we implemented central symmetry in order to reduce computer memory
requiremens by one half (we solve only for $x\ge 0$, and set ${\bf
B}(x=0^-,y,z)={\bf B}(x=0^+,-y,-z)$, and ${\bf E}(x=0^-,y,z)=-{\bf
E}(x=0^+,-y,-z)$). Another one half of computer memory is saved
when we integrate our equations in a sphere and not in a cube
centered around the neutron star. What we mean is that, in an
integration cube of size $L^3$, electromagnetic fields at the
corners remain unchanged before the spherical wave initiated by
the onset of stellar rotation (see below) reaches the surface of
the cube. There is, therefore, no need to reserve computer memory
for the cube's corners, and with proper indexing of the
rectangular grid cells, we may integrate our equations on only the
cells interior to radius $L/2$. However, PML outer boundaries can
only be implemented on planar surfaces, therefore, we had to
return to integrating in a cube.

The electric current term in eq.~(\ref{Maxwell1}) is introduced
similarly to Spitkovsky~(2006). Instead of calculating both terms
in eq.~(\ref{J}), we use only the term perpendicular to the
magnetic field (the first one) to update the electric field, and
ignore the term parallel to it. The magnetic field is updated
through eq.~(\ref{Maxwell2}). The electric field thus obtained
will have a component parallel to the magnetic field and a
component perpendicular to it. In accordance to
eq.~(\ref{EperpB}), the second component would be `killed' by the
term in the electric current that we left out. We instead `kill'
this term ourselves and thus impose the verticality condition
(eq.~\ref{EperpB}). At the same time we need to secure that the
drift velocity remains everywhere subluminal. We impose this
restriction by rescaling (after each time step) the electric field
wherever the condition $E\leq B$ is violated. The problem is that
electric and magnetic field components are defined at different
grid positions, and the above conditions are implemented using
first order interpolation between neighboring grid positions.
After each update of the electric field components the two
conditions are only approximately satisfied. We found that the
updated electric field remains almost unchanged after imposing the
verticality and rescaling conditions three times in a row. The
benefit of this approach is that it may be generalized to the
physical case where eq.~\ref{EperpB} breaks down in certain parts
of the magnetosphere when a component of the electric field
parallel to the magnetic field is allowed to develop (see \S~5).

As we said in the previous section, we insisted on the
implementation of PML outer boundaries in order to be able to run
our simulations long enough to claim with confidence that we've
reached a steady state. A detailed description of the PML
formulation can be found in Berenger~(1996), Taflove \&
Hagness~(2005) (see Appendix). The boundary layers extend beyond
the outer surfaces of our integration cube around the central
neutron star. We found that PML with 10-20 grid zones absorb very
efficiently transient electromagnetic disturbances that originate
in the integration domain without reflecting them back. Inside the
PML zone we have chosen a cubic profile for the absorption
coefficient $\sigma$. A careful choice for the maximum value of
$\sigma$ is important mostly in the cases of high values of the
inclination angle $\theta$.

\begin{figure*}[t]
\centerline{\includegraphics[angle=0,width=15cm]{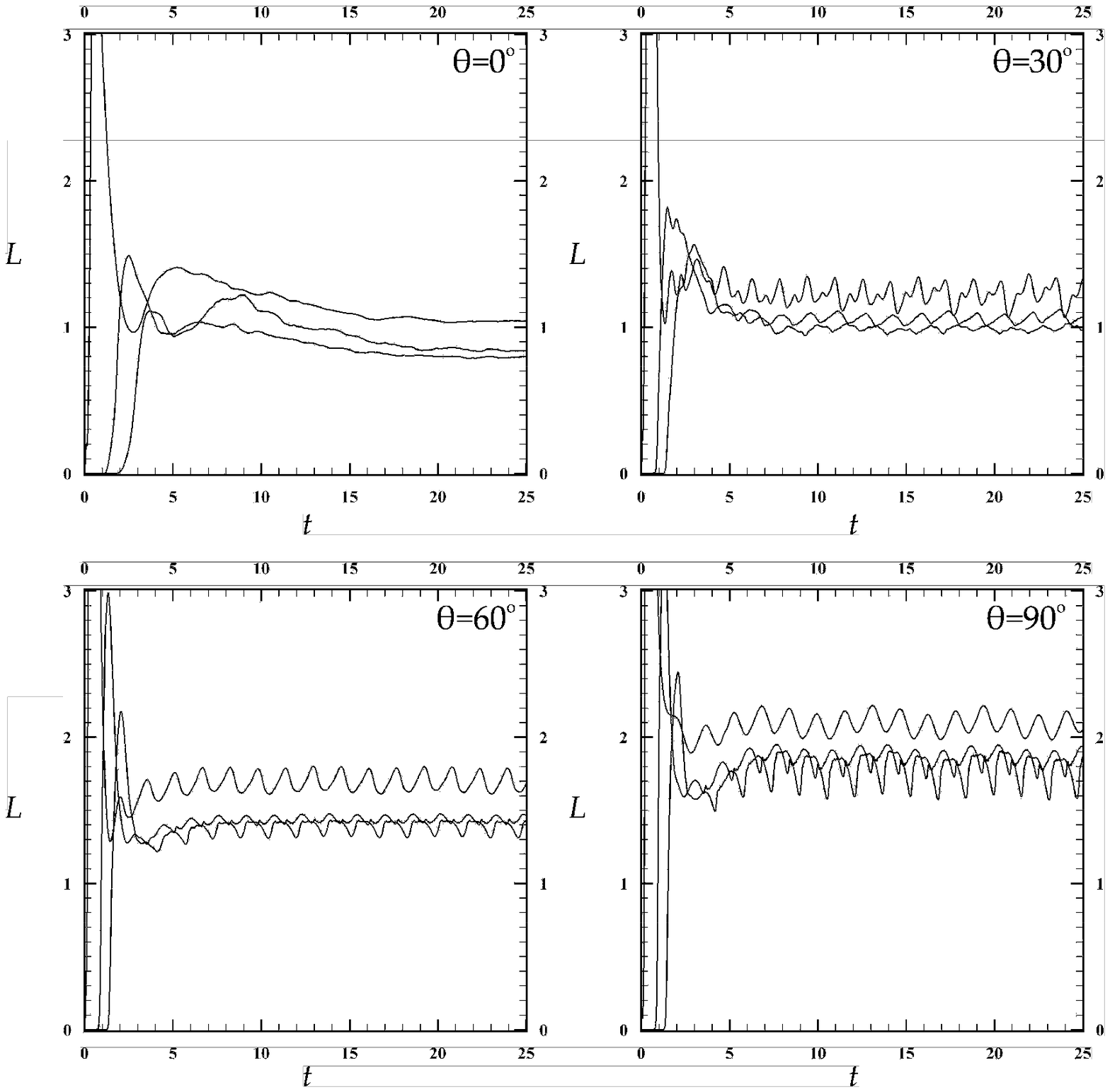}}
\caption{Total Poynting flux $L$ crossing 3 cubes with sides 0.64,
2, and 3 times $r_{lc}$ centred around the rotating neutron star
as a function of time $t$ for various values of the inclination
angle $\theta$ (top to bottom curves in each sub-panel
respectively). $L$ in units of the CKF canonical luminosity
(eq.~\ref{Spitkovsky} for $\theta=0^\circ$). $t$ in units of the
light cylinder crossing time $r_{lc}/c$ (one period of rotation is
equal to $2\pi$). The small oscillations seen at inclinations
$\theta\neq 0^o$ are an artifact of our cartesian numerical grid
and do not represent real magnetospheric oscillations (see text
for details).
} \label{fig1}
\end{figure*}

The problem that we solve is simple. We consider a spherical star
extending out to radius $r_*$. We start at $t=0$ with a dipole
magnetic field ${\bf B}_{\rm dipole}({\bf r};\theta)$ at an angle
$\theta$ with respect to some rotation axis direction. In most of
our simulations, we took the rotation axis to coincide with the
$z$-axis of our cartesian grid, but in principle, it can be
arbitrary. ${\bf E}=0$ everywhere. At $t=0^+$ we initiate the
rotation of the central star with angular velocity $\Omega_*$ as
follows: at all fixed grid points inside $r_*$ and at each time
step, we update the magnetic field with that corresponding to an
inclined dipole in rotation around the given axis direction ${\bf
B}_{\rm dipole}({\bf r},t;\theta)$, and introduce a non-zero
electric field according to eq.~\ref{corotation}. We solve the set
of equations that we presented in the previous section only on
grid points that lie outside $r_*$. Note that the smaller (larger)
$r_*$ is the more (less) physical our simulation (in a real
pulsar, $r_*/r_{lc} \leq 10^{-3}$). On the other hand, high (low)
values of $r_*$ provide a better (worse) description of the star
in a cartesian grid. We noticed that when $\theta=0^o$ the
solution is more insensitive to the chosen value of $r_*$, while
for higher values of $\theta$ lower values of $r_*$ are needed in
order to get a better description of the rotating magnetosphere.
Taking the above into account we adopted the value $r_*=0.2
r_{lc}$.


\section{A dynamic magnetosphere}

With the implementation of PML outer boundaries and central
symmetry, we were able to run our simulations for several neutron
star rotations over a cubic spatial grid centered on the neutron
star with sides 4.8 times $r_{lc}$ and spatial resolution $\delta
=0.04 r_{lc}$ on an Intel Core 2 Duo E6600 2~Gbyte RAM standard
off-the-shelf PC. Such simulations have spatial resolution
comparable to that of existing ones (Spitkovsky~2006), only now we
are able to run them for much longer times.

In Fig.~\ref{fig1} we plot the total Poynting flux calculated over
a series of cubes centered on the neutron star with sides 0.64, 2,
and 3 times $r_{lc}$ respectively as a function of time. The 3
curves are clearly displaced horizontally between them by the
amount of time it takes for the initial spherical wave induced by
the onset of the stellar rotation to cross each subsequent
calculation cube. We observe that the further away from the
central star the calculation cube is, the smaller the estimated
Poynting flux. For an ideal dissipationless calculation in steady
state, the same amount of electromagnetic energy that leaves the
star in every period of stellar rotation would cross all of the
above cubes, and the position and shape of the surfaces over which
we calculate the Poynting flux would not matter. In a real
calculation like the present one, though, some amount of
electromagnetic energy is lost between the cubes due to numerical
dissipation. In the non-axisymmetric case, this is expected to
yield periodic oscillations in the Poynting flux computed over the
above cubes at one quarter of the period of the star, which do not
represent real magnetospheric oscillations. An additional
artificial source of periodicity in the non-axisymmetric cases
comes from the fact that the representation of the rotating
stellar field on the cartesian grid repeats itself every quarter
of a period. The amplitude of these oscillations is smaller than
about $10\%$, and therefore, Fig.~1 is a test of both the
convergence and accuracy of our calculations.

In the axisymmetric case the simulation converges after about 2.5
stellar rotations ($t\geq 15$ in units of the light cylinder
crossing time $r_{lc}/c$). This justifies our insistence on
implementing PML outer boundaries. We repeated our calculation for
$\theta=30^\circ, 60^\circ$ and $90^\circ$, and at those
inclinations, the simulation converges in about 1 stellar
rotation. The final axisymmetric quasi-stationary steady state is
practically indistinguishable from the CKF steady state (see
discussion below). The further away the Poynting flux is estimated
numerically, the lower its value. Therefore, in order to estimate
the true stellar energy loss rate, we computed the Poynting flux
as close to the central star as possible, on a cube with side
equal to $0.64r_{lc}$ centred around it (that is only 3 grid zones
away from the surface of the star). Our results are consistent
with eq.~(\ref{Spitkovsky}) (within $5\%$), which constitutes an
independent confirmation of the results of Spitkovsky~(2006). The
numerical energy losses are at most on the order of 15\% at all
inclinations, and they occur mostly inside the light cylinder.

A practical problem with the general 3D case is the visualization
of the magnetic field configuration. In axisymmetry, we define
magnetic flux surfaces as surfaces of revolution of magnetic field
lines around the magnetic and rotation axis. We may thus plot the
cross sections of magnetic flux surfaces with the meridional
plane.
\begin{figure*}[t]
\centerline{\includegraphics[angle=0,width=15cm]{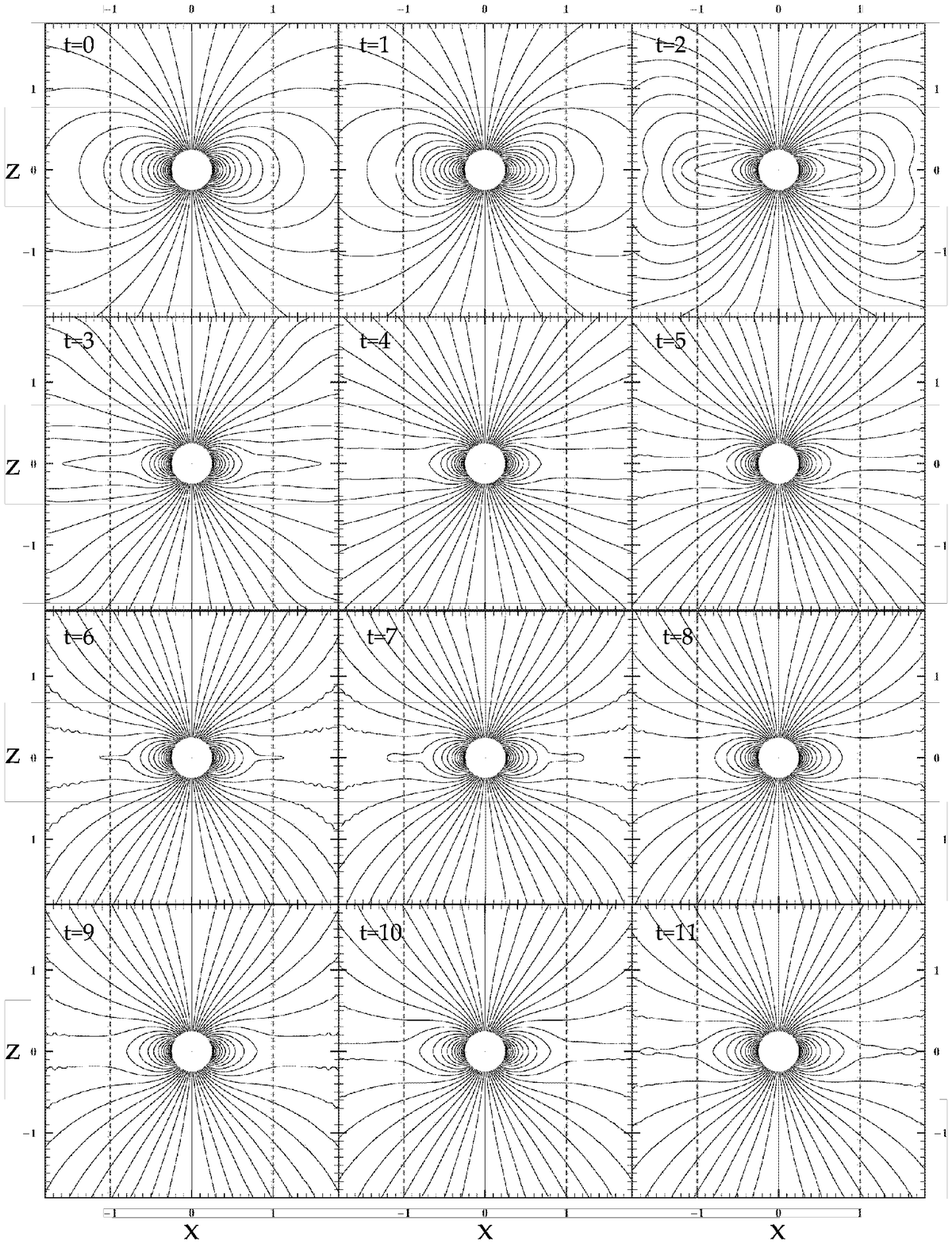}}
\caption{Time sequence of magnetic flux surfaces along the
meridional plane $(x,z)$ for an aligned rotator
($\theta=0^\circ$). Distances in units of $r_{lc}$. Times $t$ in
units of $r_{lc}/c$. Initial state at upper left. Light cylinder
shown with dashed line.} \label{fig3}
\end{figure*}
\begin{figure*}[t]
\centerline{\includegraphics[angle=0,width=15cm]{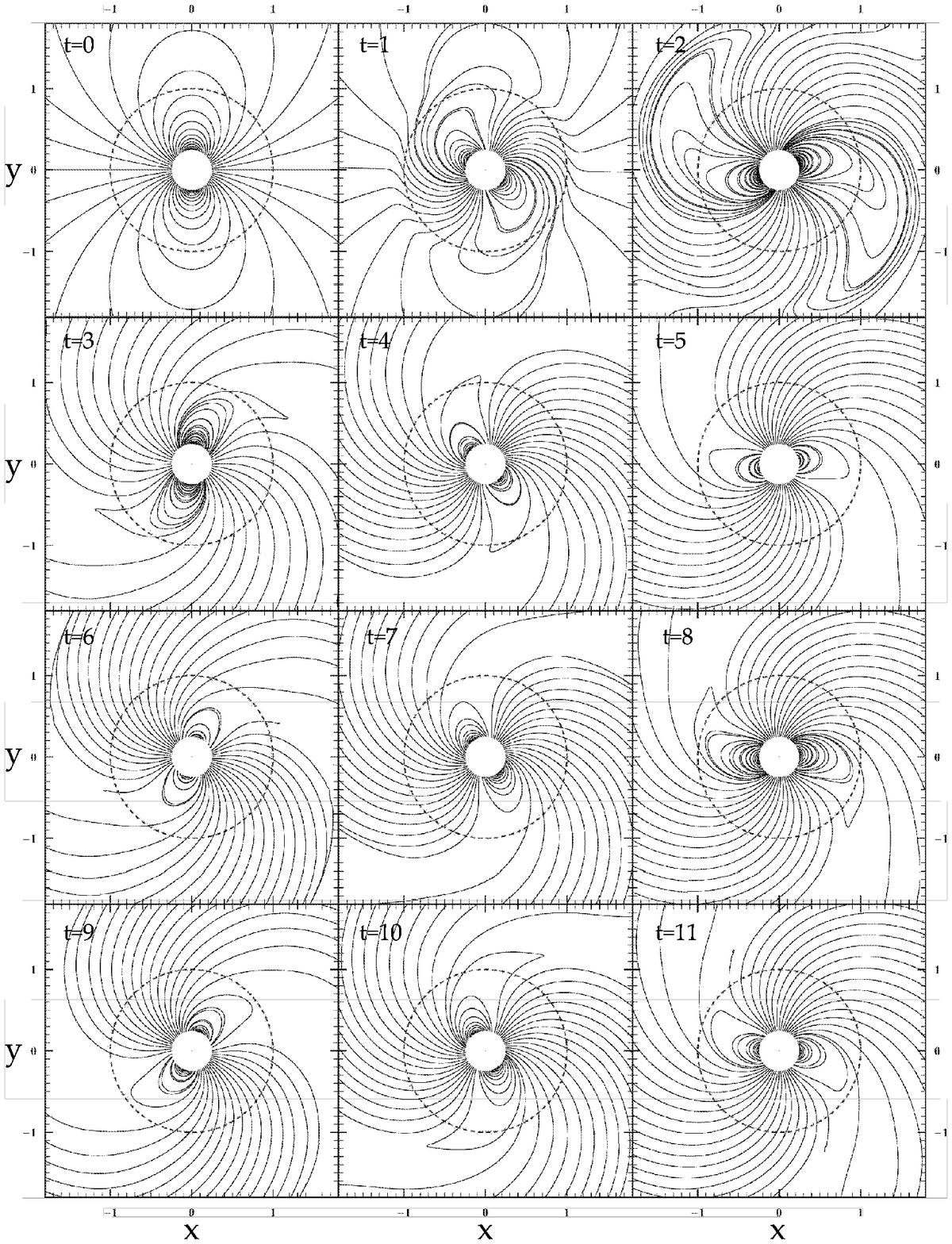}}
\caption{Time sequence of equatorial field lines originating on
the surface of the star for a $\theta=90^\circ$ oblique rotator.
Units as in Fig.~\ref{fig3}. Initial state at upper left. Light
cylinder shown with dashed line.} \label{fig4}
\end{figure*}
In the general 3D case, the choice of flux surfaces is not unique,
since any closed line along the surface of the neutron star traces
a certain flux surface along which field lines flow. When
$0^\circ<\theta<90^\circ$, the only way to visualize the solution
is by the direct drawing of 3D magnetic field lines, and plane
cuts through the magnetosphere do not mean much. When
$\theta=90^\circ$, though, magnetic field lines originating on the
equatorial plane stay on that plane (because of symmetry), and
therefore visualization of that particular case is
straightforward.

\begin{figure}[h]
\includegraphics[angle=0,scale=.80]{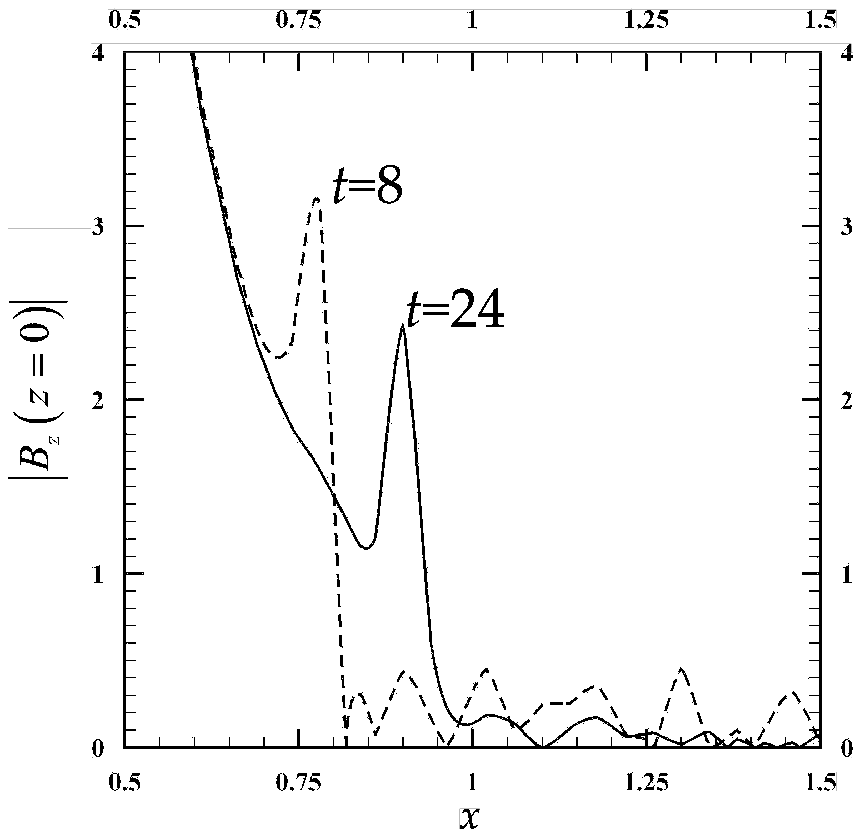}
\caption{The magnitude of the magnetic field on the equator (in
units of $B_* (r_*/r_{lc})^3/2$) at two different times during the
approach to steady state when $\theta=0^\circ$.} \label{fig2}
\end{figure}

In Fig.~\ref{fig3}, we show a time sequence of the approach to
steady state when $\theta=0^\circ$ by plotting the cuts of
axisymmetric magnetic flux surfaces with the meridional plane
$(x,z)$. The light cylinder is denoted with dashed lines. We see
that an initial wave travels out at the speed of light `informing'
the magnetosphere that the star is set in rotation at $t=0$.
Behind this wave, formerly dipolar magnetic field lines are
stretched in the radial direction, and are also twisted in the
direction opposite to the stellar rotation (`backwards'). An
equivalent way to understand this is that electric currents
develop behind this wave which carry along electric charges. These
are the ones that will populate the pulsar magnetosphere with the
space-charge density required by the steady-state solution. In
steady state, field lines that close inside the light cylinder
cannot and indeed do not contain electric currents (because of
North-South symmetry). Beyond the light cylinder, formerly closed
field lines are stretched out to infinity. It is interesting to
notice that, very quickly (within about half a rotation), a large
fraction of formerly closed field lines open up, and the closed
line region ends at about $80\%$ of the light cylinder distance
$r_{lc}$. This effect manifests itself in the evolution of the
total Poynting flux through our inner calculation cube, where
$L(0^\circ)$ reaches a value of about 1.5 times its final
steady-state value. Beyond that point, the final steady state is
gradually approached within about 2.5 stellar rotations, as the
tip of the closed line region slowly approaches the light cylinder
through a sequence of equatorial reconnection and plasmoid
generation events. Every time a plasmoid is detached, the tip and
the whole magnetosphere relax and try to readjust from the
stretching. The process repeats itself again and again, and never
actually disappears completely. The origin of this effect is
magnetic diffusivity, which in our case is entirely due to
numerical dissipation. Note that a similar behavior was observed
in the very high resolution 2D simulations of Spitkovsky~(2006)
where it took him about 20 stellar rotations to reach the final
steady state. Higher resolution simulations with adaptive mesh
refinement are needed to better capture this effect in 3D. In
Fig.~\ref{fig2} we plot the magnitude of the magnetic field on the
equator (in units of $B_* (r_*/r_{lc})^3/2$) at two different
times during the approach to steady state (compare with fig.~11 of
Timokhin~2006 and fig.~1 of Spitkovsky~2006).

After testing our code against the well understood axisymmetric
case, we proceeded with confidence to the study of nonzero values
of the inclination angle $\theta$. In Fig.~\ref{fig4} we show a
time sequence of the approach to steady state when
$\theta=90^\circ$ by plotting magnetic field lines that originate
on the surface of the star in the equatorial plane $(x,y)$. The
non-vacuum solution is different from the vacuum one in that
magnetic field lines cannot close beyond the light cylinder
(closed field lines impose corotation with the central star). As
in the axisymmetric case described above, after the passage of an
initial transient wave, the magnetosphere is populated with
electric charges due to electric currents. The approach towards
steady state is similar to the axisymmetric one but takes only
about one stellar period. Plasmoid generation from the tip of the
closed line region is also observed. Overall, our results are in
qualitative agreement with the results of Spitkovsky~(2006).

\section{Discussion}

We have shown that the implementation of PML outer boundary
conditions allows us to perform reliable time-dependent
simulations of the 3D pulsar magnetosphere on a spatial numerical
grid extending only a small distance beyond the light cylinder.
Some amount of empirical adjustment of the PML parameters is
needed. We found that a PML boundary thicker than 10 grid zones
seems to work satisfactorily. As we described above, we tested our
code against the well understood axisymmetric solution. We were
also able to reproduce the vacuum solution. We confirmed that our
conclusions are independent of the rotation axis orientation with
respect to the grid orientation by performing test runs with the
axis of rotation along various directions (e.g. along the
simulation cube diagonal $(1,1,\sqrt{2})$). The grid resolution
affects the spatial resolution and the numerical dissipation of
our simulations. We performed simulations with $d=0.08 r_{lc}$ and
obtained comparable final steady states. A smaller $d$ allows us
to implement a smaller central star. Numerical dissipation affects
the details of the time evolution. In particular, when
$\theta=0^\circ$, the approach to steady state takes around 2.5
stellar rotations when $d=0.04 r_{lc}$, and around 1 stellar
rotation when $d=0.08$. Moreover, plasmoids are generated about
once every period when $d=0.04 r_{lc}$ and about every one third
of a period when $d=0.08 r_{lc}$.

We end the present paper with a short discussion of the future
prospects for our code. A very promising avenue for research would
be to introduce physical prescriptions for the breakdown of ideal
FFE in our code. This will be implemented as follows: we will
relax the ideal MHD condition eq.~(\ref{EperpB}), and instead of
`killing' the component of the electric field that is parallel to
the magnetic field as described in \S~3 above, allow for a nonzero
parallel component as current emission models imply (e.g.
Arons~1983; Daugherty \& Harding~1982; Muslimov \& Harding~2003;
Cheng, Ho \& Ruderman~1986; Romani~1996). We will thus be able to
determine self-consistently the regions where dissipation of
electromagnetic energy takes place in the framework of a
particular high-energy emission model and thus check the validity
of that model (e.g. inner, slot, or outer gap). We will also be
able to identify the sources of plasma supply needed to make the
force-free model viable in the first place. Another interesting
feature of the pulsar phenomenon is its rich random non-periodic
time variability often referred to as `timing noise'. It is
natural for us to associate some of this variability with the
continuous plasmoid generation at the tip of the closed line
region. In a real pulsar, the situation is even more complicated
since there is no final steady state to be reached because the
light cylinder continuously moves out as the central star loses
energy and spins down. In order to address the above issues, we
are currently modifying our code to run in MPI parallel form with
a much higher spatial resolution and spatial extent.

\acknowledgements{We thank the referee, Pr. Jonathan Arons, for
his critical remarks that led to the correction of an error in the
original version of our code.}

\appendix

\section{Perfectly Matched Layer (PML) formulation}

The PML method consists of adding a PML medium around the main
computational domain where all the components of the
electromagnetic field are split into two parts. This means that
the 6 regular components $(E_x, E_y, E_z, B_x, B_y, B_z)$
integrated in the main computational domain yield 12 subcomponents
$(E_{xy}, E_{xz}, E_{yx}, E_{yz}, E_{zx}, E_{zy}, B_{xy}, B_{xz},
B_{yx}, B_{yz}, B_{zx}, B_{zy})$ inside the PML which are
integrated according to the following set of equations (Berenger
1996):
\begin{eqnarray}\label{pmlset}
\frac{\partial E_{xy}}{\partial t}+\sigma_y^e
    E_{xy} & = & \frac{\partial (B_{zx}+B_{zy})}{\partial y}\\
\frac{\partial E_{xz}}{\partial t}+\sigma_z^e
    E_{xz} & = & -\frac{\partial (B_{yz}+B_{yx})}{\partial z}\\
\frac{\partial E_{yz}}{\partial t}+\sigma_z^e
    E_{yz} & = & \frac{\partial (B_{xy}+B_{xz})}{\partial z}\\
\frac{\partial E_{yx}}{\partial t}+\sigma_x^e
    E_{yx} & = & -\frac{\partial (B_{zx}+B_{zy})}{\partial x}\\
\frac{\partial E_{zx}}{\partial t}+\sigma_x^e
    E_{zx} & = & \frac{\partial (B_{yz}+B_{yx})}{\partial x}\\
\frac{\partial E_{zy}}{\partial t}+\sigma_y^e
    E_{zy} & = & -\frac{\partial (B_{xy}+B_{xz})}{\partial y}\\
\frac{\partial B_{xy}}{\partial t}+\sigma_y^b
    B_{xy} & = & -\frac{\partial (E_{zx}+E_{zy})}{\partial y}\\
\frac{\partial B_{xz}}{\partial t}+\sigma_z^b
    B_{xz} & = & \frac{\partial (E_{yz}+E_{yx})}{\partial z}\\
\frac{\partial B_{yz}}{\partial t}+\sigma_z^b
    B_{yz} & = & -\frac{\partial (E_{xy}+E_{xz})}{\partial z}\\
\frac{\partial B_{yx}}{\partial t}+\sigma_x^b
    B_{yx} & = & \frac{\partial (E_{zx}+E_{zy})}{\partial x}\\
\frac{\partial B_{zx}}{\partial t}+\sigma_x^b
    B_{zx} & = & -\frac{\partial (E_{yz}+E_{yx})}{\partial x}\\
\frac{\partial B_{zy}}{\partial t}+\sigma_y^b
    B_{zy} & = & \frac{\partial (E_{xy}+E_{xz})}{\partial y}\ ,
\end{eqnarray}
where the conductivities $(\sigma_x^e, \sigma_y^e, \sigma_z^e)$
and $(\sigma_x^b, \sigma_y^b, \sigma_z^b)$ are defined below. When
these parameters are set to zero we obtain Maxwell's equations in
vacuum. Note that each (total) field component inside the PML is
calculated by adding its two subcomponents (e.g.
$B_x=B_{xy}+B_{xz}$). At $t=0$, each subcomponent is set equal to
one half the value of its corresponding field component.

Without loss of generality, we assume that the main computational
domain is a cube with side $L$ centered around the origin of our
cartesian coordinate system. In that case the PML technique
requires that, inside the PML,
\begin{eqnarray}\label{sigma1}
\sigma_x^a & = & 0\ \mbox{wherever}\ |x|<L/2, |y|>L/2,
    |z|>L/2 \\
\sigma_y^a & = & 0\ \mbox{wherever}\ |x|>L/2, |y|<L/2,
    |z|>L/2 \\
\sigma_z^a & = & 0\ \mbox{wherever}\ |x|>L/2, |y|>L/2,
    |z|<L/2\ ,
\end{eqnarray}
where, $a=(e,b)$. Theoretically, in the remaining PML regions the
conductivities may have constant values. In practice, the
finiteness of the grid resolution introduces artificial
reflections at the interface between the PML and the main
computational domain. For that reason it is more convenient to
consider a less discontinuous transition along the directions the
conductivities are nonzero. We have adopted the following cubic
law for the variation of the conductivities,
\begin{equation}\label{sigmacl}
    \sigma_{i}^a\approx\sigma_{max}\left(\frac{d}{D}\right)^3\ ,
\end{equation}
where, $d$ is the distance from the main cube, $D$ is the PML
thickness, and $i=(x,y,z)$. Our simulations run with $D\sim
0.4-0.8$ length units and $\sigma_{,max}\sim100-200$ inverse time
units. These values seem to work satisfactorily in both the vacuum
and non-vacuum cases.

\end{document}